\documentclass[preprint*]{JHEP3} 

\JHEPspecialurl{http://jhep.sissa.it/JOURNAL/JHEP3.tar.gz}

\usepackage{epsfig,multicol}

\newcommand\fverb{\setbox\pippobox=\hbox\bgroup\verb}
\newcommand\fverbdo{\egroup\medskip\noindent%
                        \fbox{\unhbox\pippobox}\ }
\newcommand\fverbit{\egroup\item[\fbox{\unhbox\pippobox}]}
\newbox\pippobox
\def\sk{\vspace{.4cm}}
\def\a{{\hat a}}
\def\b{{\hat b}}
\def\c{{\hat c}}
\def\d{{\hat d}}

\def\l{{\cal L}}
\def\eq{\begin{equation}}
\def\en{\end{equation}}
\def\L{{\cal L}}
\def\al{{\alpha}}
\def\be{{\beta}}
\def\ga{{\gamma}}

\def\ze{{\zeta}}
\def\vphi{{\varphi}}
\def\ddelta{\delta}

\title{Dimensional Reduction over Fuzzy Coset Spaces}

\author{P.~Aschieri$^{1,2,3,a}$, J.~Madore$^{4, b}$,
P.~Manousselis$^{5, c}$ and G.~Zoupanos$^{2,5,6, d}$\\ 
$^{1}$Dipartimento di Scienze e Tecnologie
Avanzate,\\ Universit{\'a} del Piemonte Orientale, and INFN,\\ Corso
Borsalino 54, I-15100,  Alessandria, Italy\\
$^{2}$Max-Planck-Institut f\"{u}r Physik\\ F\"{o}hringer Ring 6,
D-80805 M\"{u}nchen\\
$^{3}$Sektion Physik,
Universit\"{a}t M\"{u}nchen\\ Theresienstra{\ss} e 37, D-80333
M\"{u}nchen\\ 
$^{4}$Laboratoire de Physique
Th\'{e}orique\\Universit\'{e} de Paris-Sud, B\^{a}timent 211,
F-91405 Orsay\\  
$^{5}$Physics Department National
Technical University\\ Zografou Campus, GR-15780 Athens\\
$^{6}$Theory Division, CERN, CH-1211, Geneva 23,
Switzerland\\
E-mail: \email{aschieri@theorie.physik.uni-muenchen.de,
John.Madore@th.u-psud.fr
pman@central.ntua.gr,
George.Zoupanos@cern.ch}}
\preprint{\hepth{0310072}}

\abstract{We examine gauge theories on Minkowski
space-time times fuzzy coset spaces. This means that the extra
space dimensions instead of being a continuous coset space $S/R$
are a corresponding finite matrix approximation. The gauge theory
defined on this non-commutative setup  is reduced to four
dimensions and the rules of the corresponding dimensional
reduction are established. We investigate in particular the case
of the fuzzy sphere including the dimensional reduction of fermion
fields.}


\begin{document} 

\section{Introduction}
The theoretical efforts to establish a deeper understanding of
Nature has led to very interesting frameworks such as String
theories and Non-commutative Geometry both of which aim to
describe physics at the Planck scale. Looking for the origin of
the idea that coordinates might not commute we might have to go
back to the days of Heisenberg. In the recent years the birth of
such speculations can be found in refs.~\cite{Connes,Madore}. In
the spirit of Non-commutative Geometry also particle models with
non-commutative gauge theory were explored \cite{Connes:1990qp}
(see also \cite{Martin:1996wh}), \cite{DV.M.K., M.}. On the other
hand the present intensive research has been triggered by the
natural realization of non-commutativity of space in the string
theory context of D-branes in the presence of a constant
background antisymmetric field \cite{Connes:1997cr}. After the
work of Seiberg and Witten \cite{SW}, where a map (SW map) between
non-commutative and commutative gauge theories has been described,
there has been a lot of activity also in the construction of
non-commutative phenomenological lagrangians, for example various
non-commutative standard model like lagrangians have been proposed
\cite{Chaic, SM}\footnote{ These SM actions are mainly considered
as effective actions because they are not renormalizable. The
effective action interpretation is consistent with the SM in
\cite{SM} being anomaly free \cite{martin}. Non-commutative
phenomenology has been discussed in \cite{phenomenology}.}. In
particular in ref.~\cite{SM}, following the SW map methods
developed in refs.~\cite{Jurco:2000ja}, a non-commutative standard
model with $SU(3)\times SU(2)\times U(1)$ gauge group has been
presented. These non-commutative models represent interesting
generalizations of the SM and hint at possible new physics.
However they do not address the usual problem of the SM, the
presence of a plethora of free parameters mostly related to the ad
hoc introduction of the Higgs and Yukawa sectors in the theory. At
this stage it is worth recalling that various schemes, with the
Coset Space Dimensional Reduction (CSDR)
\cite{Forgacs:1979zs,Kapetanakis:hf,Kubyshin:vd,Bais:td} being
pioneer, were suggesting that a unification of the gauge and Higgs
sectors can be achieved in higher dimensions. Moreover the
addition of fermions in the higher-dimensional gauge theory leads
naturally after CSDR to Yukawa couplings in four dimensions. In
the successes of the CSDR scheme certainly should be added the
possibility to obtain chiral theories in four dimensions
\cite{Farakos:1986sm,Chapline:wy,Barnes:ea,Hanlon} as well as
softly broken supersymmetric or non-supersymmetric theories
starting from a supersymmetric gauge theory defined in higher
dimensions \cite{Manousselis:2001re}. \sk

In this paper we combine and exploit ideas from CSDR and
Non-commutative Geometry. We consider the dimensional reduction of
gauge theories defined in high dimensions where the internal space
is a fuzzy space (matrix manifold). In the  CSDR one assumes that
the form of space-time is $M^{D}=M^{4} \times S/R$ with $S/R$ a
homogeneous space (obtained as the quotient of the Lie group $S$ via the
Lie subgroup $R$).  Then a gauge theory with gauge group $G$ 
defined on $M^{D}$ can be dimensionally reduced to $M^{4}$ in an
elegant way using the symmetries of $S/R$,
in particular the resulting
four dimensional gauge group is a subgroup of $G$. 
In the present work we
will apply the method of CSDR in the case where the internal part
of the space-time is a finite approximation of the homogeneous
space $S/R$, i.e. a fuzzy coset (fuzzy cosets are studied in
\cite{Grosse:1999ci,Trivedi,Bala,Balachandran:2001dd}). In particular
we study the fuzzy sphere case \cite{Mad}. Fuzzy spaces are
obtained by deforming the algebra of functions on their
commutative parent spaces. The algebra of functions (from the
fuzzy space to complex numbers) becomes finite dimensional and
non-commutative, indeed it becomes a matrix algebra. Therefore,
instead of considering the algebra of functions $Fun(M^{D})\sim
Fun(M^{4}) \times Fun(S/R)$ we consider the algebra $A= Fun(M^{4})
\times M_{n}$ where $Fun(M^{4})$ is the usual commutative algebra
of functions on Minkowski space $M^{4}$ and $M_{n}$ is the finite
dimensional non-commutative algebra of matrices that approximates
the coset; on this finite dimensional algebra we still have the
action of the symmetry group $S$. This very property will allow us
to apply the CSDR scheme to fuzzy cosets.
In the parent theory on $M^D=M^4\times (S/R)_F$ the
non-commutativity will lead us to consider the gauge groups
$G=U(1)$ and more generally $G=U(P)\,$\footnote{Alternatively one
could also formulate the
  non-commutativity of $(S/R)_F$
in terms of a star product. Then, using SW map, it is possible as
in \cite{Jurco:2000ja} to consider arbitrary gauge groups $G$.
This approach relies on a perturbative expansion in the
non-commutativity parameter
$\theta^{\a\b}(\hat{X})=[\hat{X}^\a,\hat{X}^\b]$ and therefore,
see e.g. (\ref{vt}), is particularly promising when the fuzzy
manifold is described by $n\times n$ matrices in the limit
$n\to \infty$.}. 
Notice that there is no a priori relation between the
gauge group  $G=U(P)$ and the groups $S$ and $R$.
\sk

In summary, gauge theories have been studied on non-commutative
Minkowski space as well as on the product space commutative
Minkowski times internal non-commutative space
\cite{Connes:1990qp, Martin:1996wh, DV.M.K., M.}, see also ref.
\cite{Cast} and ref. \cite{Aschieri:2002vn}, where the internal
space is the lattice of a finite group (and non-commutative
geometry techniques allow to describe this lattice as a manifold
\cite{Dimakis:1994qq}). CSDR is a unification scheme for obtaining
realistic particle models, and the study of CSDR in the
non-commutative context provides new particle models that might be
phenomenologically relevant. One could study CSDR with the whole
parent space $M^D$ being non-commutative or with just
non-commutative Minkowski space or non-commutative internal space.
We specialize here to this last situation and thus in the end
obtain Lorentz covariant theories on commutative Minkowski space.
We further specialize to fuzzy non-commutativity, i.e. to matrix
kind non-commutativity. We thus consider non-commutative spaces
like those studied in refs. \cite{DV.M.K., M.} and implementing
the CSDR principle on these spaces we obtain new
particle models. \sk

The paper is organized as follows. We first recall the geometry of
fuzzy coset spaces, with the leading example of the fuzzy sphere.
In particular we study the Lie derivative on spinors.
Non-commutative  gauge fields and non-commutative gauge
transformations are then also recalled. 
In Section 3 we briefly review the CSDR scheme in the commutative case 
and then implement the CSDR principle on fuzzy cosets, 
we thus obtain a set of contraints --namely the CSDR constraints-- 
that the gauge and matter fields must satisfy. 
Next we first reinterpret an action on $M^4\times (S/R)_F$
with $G=U(P)$ gauge group as an action  on $M^4$
with $U(nP)$ gauge group.
We then impose and solve the CSDR constraints and obtain
the gauge group and the particle content of the reduced
four-dimensional action. 
Discussions and conclusions are in Section 4.

\section{Fuzzy sphere and fuzzy coset spaces geometry}\label{2}

In this section first we describe the fuzzy sphere and study
spinor fields on the fuzzy sphere, then we briefly present more
general fuzzy coset spaces. For the definition of the fuzzy sphere
and the gauge theory over the fuzzy sphere  we follow ref.
\cite{Mad} (see also ref. \cite{Madore:1991bw}).
 A fuzzy manifold is a discrete matrix
approximation to a continuous manifold. The approximation is such
that the discretized space preserves its continuum symmetries
\cite{Balachandran:2001dd}, a fact that will allow us to apply the
CSDR. A method in order to discretize a manifold is to single out
a (finite) subspace of the space of functions on the manifold. One
would also like this subspace to be invariant under
multiplication. As a simple example consider the Fourier analysis
of a function on a circle,
\begin{equation}
f(\theta) = \sum_{n= - \infty}^{\infty} f_{n} e^{in\theta}.
\end{equation}
A discretized version of the circle can be achieved replacing the
algebra of functions on the circle with the space of functions
that do not exceed a given frequency $N$. We then write
\begin{equation}
f_{N}(\theta) = \sum_{n= - N}^{N} f_{n} e^{in\theta}.
\end{equation}
for a generic function, $f_{N}(\theta)$ being an approximation of
$f(\theta)$. However the product of two such functions will in
general extend to frequencies up to $2N$ and so the space of
truncated functions does not close under multiplication, we cannot
speak of an algebra of truncated functions. The same is true for
the harmonic analysis on the sphere or any other coset $S/R$. The
solution is in the definition of a non-commutative product (a
matrix product) such that the space of truncated functions closes
under this new  product.

\sk

The algebra of functions on the ordinary sphere can be generated
by the coordinates of {\bf{R}}$^3$ modulo the relation $
\sum_{\hat{a}=1}^{3} {x}_{\hat{a}}{x}_{\hat{a}} =r^{2}$. The fuzzy
sphere $S^2_{F}$ at fuzziness level $N$ is the non-commutative
manifold whose coordinate functions $ \hat{X}_{\hat{a}}
=\hat{X}^{\hat{a}}$ are $(N+1) \times (N+1)$ hermitian matrices
proportional to the generators of the $(N+1)$-dimensional
representation of $SU(2)$, $\hat{X}_{\hat{a}}=\kappa J^{\hat{a}}$.
They satisfy the condition $ \sum_{\hat{a}=1}^{3}
\hat{X}_{\hat{a}} \hat{X}_{\hat{a}} =r^{2}$ and the following
commutation relations
\begin{equation}
\label{vt} [\hat{X}_{\hat{a}}, \hat{X}_{\hat{b}}] =i\kappa
C_{\hat{a} \hat{b} \hat{c}} \hat{X}_{\hat{c}},
\end{equation}
where
$ \kappa = \lambda_{N}
r$ with  $ \lambda_{N} =
1/\scriptstyle{{\sqrt{\frac{N}{2}(\frac{N}{2}
      +1)}}}\,$
[we use $J^2={\frac{N}{2}(\frac{N}{2} +1)}$ for the $N+1$
dimensional irrep. of $SU(2)$]. If we define 
\eq
 X_{\hat{a}} = \frac{1}{i\kappa r}
\hat{X}_{\hat{a}}=\frac{1}{i r} J_\a\label{2.4}
\en
we have
\begin{equation}
[ X_{\hat{a}}, X_{\hat{b}} ] = C_{\hat{a} \hat{b} \hat{c}}
X_{\hat{c}}\label{2.5}
\end{equation}
with
 $C_{\hat{a} \hat{b} \hat{c}} = {\epsilon_{\hat{a} \hat{b}
\hat{c\,}} }{/ _{{{{\mbox{$ r$}}}}}} $ and  
 $$ \sum_{\hat{a} =1}^{3}X_{\hat{a}} X_{\hat{a}}
= - \frac{\lambda_{N}^{-2}}{r^2}.$$ 
In order to
describe the algebra of the fuzzy sphere $S^2_F$ we can
equivalently use the $\hat{X}_{\hat{a}}$ or the $X_{\hat{a}}$
generators; in the following we will work in the latter basis.

A function on the fuzzy sphere is a symmetric polynomial in the
$X^\a$ coordinates. Since these coordinates are proportional to
the $N+1$ dimensional irrep. of $SU(2)$ we have that any
polynomial in the $X^\a$ can be rewritten as a symmetric
polynomial of degree $\leq N$, and any $(N+1)\times (N+1)$ matrix
can be expanded as a symmetric  polynomial in the $X^\a$. Thus the
space of functions on the fuzzy sphere $S^2_F$ at level $N$ has
dimension $(N+1)^2$. A convenient basis for this space is provided
by the constant function $\mathbf{1}$ (the identity matrix) plus
the non-commutative spherical harmonics up to level $N$
\begin{equation}
\hat{Y}_{lm} = r^{-l} \sum_{\a}f^{(lm)}_{\a_{1}, \a_{2}, \ldots,
\a_{l}}X^{{\a_{1}}} \ldots X^{{\a_{l}}}~~,~~~~~~~~l\leq N
\end{equation}
with $f^{lm}_{\a_{1}, \a_{2}, \ldots, \a_{l}}$ the traceless and
symmetric tensor of the ordinary spherical harmonics. Finally a
generic function on the fuzzy sphere takes the form
\begin{equation}
f = \sum_{l=0}^{N} \sum_{m=-l}^{l} f_{lm} \hat{Y}_{lm} \ , \
\
\end{equation}
i.e. corresponds to an ordinary function on the commutative sphere
with a cutoff on the angular momentum. Obviously this space of
truncated functions  is closed under the non-commutative
$(N+1)\times (N+1)$ matrix product.

\sk On the fuzzy sphere there is a natural $SU(2)$ covariant
differential calculus. This calculus is three dimensional; the
fact that the tangent space  to the fuzzy sphere is three and not
two dimensional is a typical aspect of non-commutative spaces. The
three derivations $e_\a$ along $X_\a$ of a function $ f$ are given
by
\begin{equation}
e_\a({f})=[X_{\hat{a}}, {f}]\,.\label{derivations}
\end{equation}
Accordingly the action of the Lie derivatives on functions is
given by
\begin{equation}\label{LDA}
{\cal L}_{\hat{a}} f = [{X}_{\hat{a}},f ]~,
\end{equation}
they satisfy the Leibniz rule and the $SU(2)$ Lie algebra relation
\begin{equation}\label{LDCR}
[ {\cal L}_{\hat{a}}, {\cal L}_{\hat{b}} ] = C_{\hat{a} \hat{b}
\hat{c}} {\cal L}_{\hat{c}}.
\end{equation}
In the $N \to \infty$ limit the derivations $e_\a$ become
\begin{equation}
e_{\hat{a}} = C_{\hat{a} \hat{b} \hat{c}} x^{\hat{b}}
\partial^{\hat{c}}\,
\end{equation}
and only in this commutative limit the tangent space becomes two
dimensional. The exterior derivative is given by
\begin{equation}
d{f} = [X_\a,{f}]\theta^{\hat{a}}
\end{equation}
with $\theta^{\hat{a}}$ the one-forms dual to the vector fields
$e_{\hat{a}}$, $<e_\a,\theta^\b>=\delta_\a^\b$. The space of
one-forms is generated by the $\theta^\a$'s in the sense that for
any one-form $\omega=\sum_i f_i (d h_i)\,t_i$ we can always write
$\omega=\sum_{\a=1}^3{\omega}_\a\theta^\a$ with given functions
${\omega}_\a$ depending on the functions ${f}_i$,  ${h}_i$ and
${t}_i$. {}From $0={\cal L}_\a <e_\b,\theta^\c>=<{\cal L}_\a
e_\b,\theta^\c>+ <e_\b,{\cal L}_\a\theta^\c>$ and ${\cal L}_\a
(e_\b)=C_{\a\b\c}e_\c$ [cf. (\ref{LDCR})] we obtain the action of
the Lie derivatives on one-forms,
\begin{equation}\label{2.16}
{\cal L}_{\hat{a}}(\theta^{\hat{b}}) =  C_{\a\b\c}
\theta^{\hat{c}}.
\end{equation}
It is then easy to check that the Lie derivative commutes with the
exterior differential $d$, i.e. $SU(2)$ invariance of the exterior
differential. On a general one-form
$\omega=\omega_{\hat{a}}\theta^{\hat{a}}$ we have
\begin{eqnarray}
{\cal L}_{\hat{b}}\omega={\cal
L}_{\hat{b}}(\omega_{\hat{a}}\theta^{\hat{a}})&=&({\cal
L}_{\hat{b}} \omega_{\hat{a}})\theta^{\hat{a}}-
\omega_{\hat{a}}C^{\hat{a}}_{\
\hat{b}\hat{c}}\theta^{\hat{c}}\nonumber\\
&=&\left[X_{\hat{b}},\omega_{\hat{a}}\right]\theta^{\hat{a}}-\omega_{\hat{a}}C^{\hat{a}}_{\
\hat{b} \hat{c}}\theta^{\hat{c}}
\end{eqnarray}
and therefore
\begin{equation}
({\cal
L}_{\hat{b}}\omega)_{\hat{a}}=\left[X_{\hat{b}},\omega_{\hat{a}}\right]-
\omega_{\hat{c}}C^{\hat{c}}_{\ \hat{b}  \hat{a}}~;\label{fund}
\end{equation}
this formula will be fundamental for formulating the CSDR
principle on fuzzy cosets. Similarly, from ${\l}_\b(v)={\l}_\b(v^\a
e_\a)=[X_\b,v^\a]+{\l}_\b(e_\a)$ we have \eq ({\l}_\b
v)^\a=\left[X_{\hat{b}},v^{\hat{a}}\right]- v_{\hat{c}}C_{\hat{c}
\hat{b}  \hat{a}}~.\label{lievect}
\en
The differential geometry on  the product space Minkowski times
fuzzy sphere, $M^{4} \times S^2_{F}$, is easily obtained from that
on $M^4$ and on $S^2_F$. For example a one-form $A$ defined on
$M^{4} \times S^2_{F}$ is written as
\begin{equation}\label{oneform}
A= A_{\mu} dx^{\mu} + A_{\hat{a}} \theta^{\hat{a}}
\end{equation}
with $A_{\mu} =A_{\mu}(x^{\mu}, X_{\hat{a}} )$ and $A_{\a}
=A_{\a}(x^{\mu}, X_{\hat{a}} )$.

\sk There are different approaches to the study of spinor fields
on the fuzzy sphere \cite{Grosse:1995jt, Carow-Watamura:1996wg}.
Here we follow ref. \cite{Madore} (section 8.2)\footnote{For a
discussion of chiral fermions and index theorems on matrix
approximations of manifolds see ref. \cite{Dolan:2002af}. }. In
the case of the product of Minkowski space and the fuzzy sphere,
$M^{4} \times S^2_{F}$, we have seen that the geometry resembles
in some aspects ordinary commutative geometry in seven dimensions.
As $N \to \infty$ it returns to the ordinary
six-dimensional geometry. Let $g_{AB}$ be the Minkowski metric in
seven dimensions and $\Gamma^{A}$ the associated Dirac matrices
which can be in the form \eq \Gamma^{A} =
(\Gamma^\mu,\Gamma^\a)=(1 \otimes \gamma^{\mu}, \sigma^{\a}
\otimes \gamma_{5}). \label{Gamma7}
\en
The space of spinors must be a left module with respect to the
Clifford algebra. It is therefore a space of functions with values
in a vector space ${\cal H}'$ of the form $$ {\cal H}' = {\cal H}
\otimes C^{2} \otimes C^{4}, $$ where ${\cal H} $ is an $M_{N+1}$
module. The geometry resembles but is not really
seven-dimensional, e.g. chirality can be defined and the fuzzy
sphere admits chiral spinors. Therefore the space $H'$ can be
decomposed into two subspaces ${\cal H}'_{\pm} = \frac{1 \pm
\Gamma}{2}{\cal H}'$, where $\Gamma$ is the chirality operator of
the fuzzy sphere \cite{Madore,Ydri:2001pv}. The same holds for
other fuzzy cosets such as $(SU(3)/U(1) \times U(1))_{F}$
\cite{Trivedi}.

In order to define the action of the Lie derivative ${\cal{L}}_\a$
on a spinor field $\Psi$, we write
\begin{equation}
\Psi=\ze_\al\psi_\al,
\end{equation}
where $\psi_\al$ are the components of $\Psi$  in the $\ze_\al$
basis. Under a spinor rotation $\psi_\al\to
S_{\al\be}\psi_\be$ the bilinear $\bar\psi \Gamma^\a\psi$
transforms as a vector $v^\a\to \Lambda_{\a\b}v^\b$. The
Lie derivative on the basis $\ze_\al$ is given by
\begin{equation}\label{spinlie1}
{\cal L}_{\hat{a}} \ze_\al =\ze_\be\tau^\a_{\be\al},
\end{equation}
where \eq \tau^\a=\frac{1}{2}C_{\hat{a} \hat{b} \hat{c}}
\Gamma^{\hat{b}
  \hat{c}}~~,~~~~~~ \Gamma^{\hat{b} \hat{c} }
= -{1\over 4}(\Gamma^{\hat{b}} \Gamma^{\hat{c}} - \Gamma^{\hat{c}}
\Gamma^{\hat{b}})~. \label{definitions}
\en
Using that  $\Gamma^{\b\c}$ are a rep. of the orthogonal algebra
and then using the Jacoby identities for $C_{\a\b\c}$ one has
$
[\tau^\a,\tau^\b]=C_{\a\b\c}\tau^\c
$
from which it follows that the Lie derivative on spinors gives a
representation of the Lie algebra,
 \eq
[ {\cal L}_{\hat{a}}, {\cal L}_{\hat{b}} ] \ze_\al= C_{\hat{a}
\hat{b} \hat{c}}\, {\cal L}_{\hat{c}}\,\ze_\al~~.
\en
On a generic spinor $\Psi$, applying the Leibniz rule  we have \eq
{\L}_\a\Psi=
\ze_\al[X_\a,\psi_\al]+\ze_\be \tau^\a_{\be\ga}\psi_\ga
\en
and of course $ [  {\cal L}_{\hat{a}}, {\cal L}_{\hat{b}} ] \Psi=
C_{\hat{a} \hat{b} \hat{c}}\, {\cal L}_{\hat{c}}\,\Psi$; we also
write \eq \delta_\a\psi_\al=({\L}_\a\Psi)_\al=
[X_\a,\psi_\al]+\tau^\a_{\al\ga}\psi_\ga~.\label{spinlie11}
\en
The action of the Lie derivative ${\cal{L}}_\a$ on the adjoint
spinor is obtained considering the adjoint of the above
expression, since $(X_\a)^\dagger=-X_\a$,
$(\tau^\a)^\dagger=-\tau^\a$,  $[\tau^\a,\Gamma_0]=0$ we have
\begin{equation}\label{spinlie2}
\delta_\a\bar\psi_\al=  [X_{\hat{a}},{\bar\psi_\al}] -
\bar\psi_\ga\tau^\a_{\ga\al}.
\end{equation}
One can then check that the variations (\ref{spinlie11}) and
(\ref{spinlie2}) are consistent with $\psi^\dagger\Gamma^0\psi$
being a scalar. Finally we have compatibility among the Lie
derivatives (\ref{spinlie11}), (\ref{spinlie2}) and (\ref{fund}):
$$\delta_\a(\bar\psi\Gamma^\d\psi)=(\delta_\a\bar\psi)\Gamma^\d\psi+
\bar\psi\Gamma^\d\delta_\a\psi=[X_\a,\bar\psi\Gamma^\d\psi]
+\bar\psi[\Gamma^\d,\tau^\a]\psi= [X_\a,\bar\psi\Gamma^\d\psi] -
C_{\a\d\c}\bar\psi\Gamma^\c\psi $$ (and
$\delta_\a(\bar\psi\Gamma^\mu\psi)=[X_\a,\bar\psi\Gamma^\mu\psi]$).
This immediately generalizes to higher tensors ${\bar
\psi}\,\Gamma_{\!\d_1}\ldots\Gamma_{\!\d_i} \psi$.

\sk

The sphere $S^{2}$ is the complex projective space $CP^{1}$ . The
generalization of the fuzzy sphere construction to $CP^{2}$ and
its $spin^{c}$ structure was given in ref.~\cite{Grosse:1999ci},
whereas the generalization to $CP^{M-1}= SU(M)/U(M-1)$ and to
Grassmannian cosets was given in ref.~\cite{Balachandran:2001dd}.
While a set of coordinates on the sphere is given by the {\bf
R}$^3$ coordinates $x^\a$ modulo the relation $\sum_\a x^\a
x^\a=r^2$, a set of coordinates on $CP^{M-1}$ is given by $x^\a$,
$\a=1,\ldots M^2-1$ modulo the relations 
\eq
 x^{\hat{a}}x_{\hat{a}}
= \frac{2(M-1)}{M}r^2 \ , \ \ \ d_{\hat{a}\hat{b}}^{\ \ \ \hat{c}}
x^{\hat{a}} x^{\hat{b}} = \frac{{2}(M-2)}{M}r x^{\hat{c}}\,, 
\en
where $d_{\hat{a}\hat{b}}^{\ \ \ \hat{c}}$ are the components of
the symmetric invariant tensor of $SU(M)$. Then $CP^{M-1}$ is
approximated, at fuzziness level $N$, by $n\times n$ dimensional
matrices $X_{\hat{a}}$, $\hat{a}= 1, \ldots, M^{2}-1$. These are
proportional to the generators $J_\a$ of $SU(M)$ considered in the
$n={{(M-1+N)!} \over{(M-1)!N!}}$ dimensional irrep., obtained from
the $N$-fold symmetric tensor product of the fundamental
$M$-dimensional representation of $SU(M)$. As in (\ref{2.4}) we set 
$X_\a=\frac{1}{i r}J_\a$ so that  
\eq 
\sum_{\hat{a}=1}^{3}
{X}_{\hat{a}} {X}_{\hat{a}} = -\frac{C_n}{r^2}\label{cas2}
{}~~~~,~~~~~[ X_{\hat{a}}, X_{\hat{b}} ] = C_{\hat{a} \hat{b} \hat{c}}
X_{\hat{c}}
\en
where $C_n$
is the quadratic casimir of the given
$n$-dimensional irrep.,  and $r C_{\a\b\c}$ are now 
the $SU(M)$ structure constants.
More generally \cite{Trivedi} we consider 
fuzzy coset spaces $(S/R)_F$ described by non-commuting
coordinates $X_\a$ that are proportional to the generators of a given
$n$-dimensional irrep. of the compact Lie group $S$ and thus in 
particular satisfy the conditions (\ref{cas2}) where now 
$r C_{\a\b\c}$ are the $S$ structure constants (the extra constraints
associated with the given $n$-dimensional irrep. 
determine the subgroup $R$ of $S$ in $S/R$).
The differential calculus on these fuzzy spaces can be constructed 
as in the case for the fuzzy sphere. For example there are $dim S$ 
Lie derivatives, they are given by eq. (\ref{LDA}) and satisfy the
relation (\ref{LDCR}). On these fuzzy spaces we consider 
the space of spinors to be a left module with respect to 
the Clifford algebra given by (\ref{Gamma7}), where now 
the $\sigma^\a$'s are 
replaced by the $\gamma^\a$'s, the gamma matrices on $R^{dim S}$; 
in particular all the formulae concerning Lie derivatives on spinors 
remain unchanged.

\subsection{Non-commutative gauge fields and transformations}
Gauge fields arise in non-commutative geometry and in particular
on fuzzy spaces very naturally; they are linked to the notion of
covariant coordinate \cite{Madore:2000en}. Consider a field
$\phi(X^\a)$ on a fuzzy space described by the non-commuting
coordinates $X^\a$. An infinitesimal gauge transformation
$\ddelta\phi$ of the field $\phi$ with gauge transformation
parameter $\lambda(X^\a)$ is defined by
\begin{equation}
\ddelta\phi(X) = \lambda(X)\phi(X). 
\end{equation}
This is an infinitesimal abelian $U(1)$ gauge transformation if
$\lambda(X)$ is just an antihermitian function of the coordinates
$X^\a$, it is an infinitesimal nonabelian $U(P)$ gauge
transformation if $\lambda(X)$ is valued in ${\rm{Lie}}(U(P))$,
the Lie algebra of hermitian $P\times P$ matrices; in the
following we will always assume ${\rm{Lie}}(U(P))$ elements to
commute with the coordinates $X^\a$. The coordinates $X$ are
invariant under a gauge transformation
\begin{equation}
\ddelta X_{\hat{a}} = 0~;
\end{equation}
multiplication of a field on the left by a coordinate is then not
a covariant operation in the non-commutative case. That is
\begin{equation}
\ddelta(X_{\hat{a}}\phi) = X_{\hat{a}}\lambda(X)\phi,
\end{equation}
and in general the right hand side is not equal to
$\lambda(X)X_{a}\phi$. Following the ideas of ordinary gauge
theory one then introduces covariant coordinates $\vphi_{\hat{a}}$
such that
\begin{equation}\label{covcoord}
\ddelta(\vphi_{\hat{a}}{}\phi) = \lambda\vphi_{\hat{a}}{}\phi~,
\end{equation}
this happens if
\begin{equation}
\label{covtr} \ddelta(\vphi_{\hat{a}})=[\lambda,\vphi_{\hat{a}}]~.
\end{equation}
We also set
\begin{equation}
\vphi_{\hat{a}} \equiv X_{\hat{a}} + A_{\hat{a}}~
\end{equation}
and interpret $A_\a$ as the gauge potential of the non-commutative
theory; then $\vphi_\a$ is the non-commutative analogue of a
covariant derivative. The transformation properties of
$A_{\hat{a}}$ support the interpretation of $A_\a$ as gauge field;
they arise from requirement (\ref{covtr}),
\begin{equation}
\ddelta A_{\hat{a}} = -[ X_{\hat{a}}, \lambda] +
[\lambda,A_{\hat{a}}]~.
\end{equation}
Correspondingly we can define a tensor $F_{\hat{a} \hat{b}\,}$,
the analogue of the field strength, as
\begin{eqnarray}
\label{2.33} F_{\hat{a} \hat{b}} &=& [ X_{\hat{a}}, A_{\hat{b}}] -
[ X_{\hat{b}}, A_{\hat{a}} ] + [A_{\hat{a}} , A_{\hat{b}} ] -
C^{\hat{c}}_{\ \hat{a} \hat{b}}A_{\hat{c}}\\&=& [ \vphi_{\hat{a}},
\vphi_{\hat{b}}] - C^{\hat{c}}_{\ \hat{a} \hat{b}}\vphi_{\hat{c}}.
\end{eqnarray}
This tensor transforms covariantly
\begin{equation}
\ddelta F_{\hat{a} \hat{b}} = [\lambda, F_{\hat{a} \hat{b}}]~.
\end{equation}
Similarly, for a spinor $\psi$ in the adjoint representation, the
infinitesimal gauge transformation is given by
\begin{equation}
\ddelta \psi = [\lambda, \psi]~,
\end{equation}
while for a spinor in the fundamental the infinitesimal gauge
transformation is given by \eq \delta\psi=\lambda\psi~.
\en

\section{Coset Space Dimensional Reduction (CSDR)}

{}First we briefly recall the CSDR scheme in the commutative case.
It is indeed instructive to compare the commutative and the fuzzy
case. The latter is described in the next subsection which is self
contained.

One way to dimensionally reduce a gauge theory on $M^{4} \times
S/R$ with gauge group $G$ to a gauge theory on $M^4$, is to
consider field configurations that are invariant under $S/R$
transformations. Since the action of the group $S$ on the coset
space $S/R$ is transitive (i.e., connects all points), we can
equivalently require the fields in the theory to be invariant
under the action of $S$ on $S/R$. Infinitesimally, if we denote by
$\zeta_{\a}$ the Killing vectors on $S/R$ associated to the
generators $T^\a$ of $S$, we require the fields to have zero Lie
derivative along $\zeta_{\a}$. For scalar fields this is
equivalent to requiring independence under the $S/R$ coordinates.
The CSDR scheme dimensionally reduces a gauge theory on $M^{4}
\times S/R$ with gauge group $G$ to a gauge theory on $M^4$
imposing a milder constraint, namely the fields are required to be
invariant under the $S$ action up to a $G$ gauge transformation
\cite{Forgacs:1979zs, Kapetanakis:hf, Kubyshin:vd}. Thus we have,
respectively for scalar fields $\phi$ and the one-form gauge field
$A$ \eq {\L}_{\zeta_\a}\phi=\delta^{W_\a}\phi=W_\a\phi~~,
\en
\eq {\L}_{\zeta_\a}A=\delta^{W_\a}A=-DW_\a \label{condCSDR}~,
\en
where $\delta^{W_\a}$ is the infinitesimal gauge transformation
relative to the gauge parameter ${W_\a}$ that depends on the coset
coordinates (in our notations  $A$ and $W_\a$  are 
antihermitian and the covariant derivative reads $D=d+A$).
The gauge parameters ${W_\a}$ obey a consistency condition which
follows from the relation
\begin{equation}\label{cc}
[{\cal L}_{\xi_{\hat{a}}},{\cal L}_{\xi_{\hat{b}}}] = {\cal
L}_{[\xi_{\hat{a}},\xi_{\hat{b}}]}
\end{equation}
and transform under a gauge transformation $\phi\to g\phi$
as
\begin{equation}\label{W}
W_{\hat{a}} \to gW_{\hat{a}}g^{-1} +
({\L}_{\xi_{\hat{a}}}g)g^{-1}.
\end{equation}
Since two points of the coset are connected by an
$S$-transformation which is equivalent to a gauge transformation,
and since the Lagrangian is gauge invariant, we can study the
above equations just at one point of the coset, let's say $y^a=0$,
where we denote by $(x^\mu,y^a)$ the coordinates of $M^4\times
S/R$, and we use $ \a,a,i$ to denote $S,\,S/R$ and $R$ indices. In
general, using (\ref{W}),  not all the $W_{\hat{a}}$ can be gauged
transformed to zero at $y^a=0$, however one can choose $W_{a} =0$
denoting by $W_{i}$ the remaining ones. Then the consistency
condition which follows from eq.~(\ref{cc}) implies that $W_{i}$
are constant and equal to the generators of the embedding of $R$
in $G$ (thus in particular $R$ must be embeddable in $G$; we write
$R_G$ for the image of $R$ in $G$).

The detailed analysis of the constraints given in
refs.~\cite{Forgacs:1979zs, Kapetanakis:hf} provides us with the
four-dimensional unconstrained fields as well as with the gauge
invariance that remains in the theory after dimensional reduction.
Here we give the results. The components $A_{\mu}(x,y)$ of the
initial gauge field $A_{M}(x,y)$ become, after dimensional
reduction, the four-dimensional gauge fields and furthermore they
are independent of $y$. In addition one can find that they have to
commute with the elements of the $R_{G}$ subgroup of $G$. Thus the
four-dimensional gauge group $H$ is the centralizer of $R$ in $G$,
$H=C_{G}(R_{G})$. Similarly, the $A_a(x,y)$ components of
$A_{M}(x,y)$ denoted by $\phi_a(x,y)$ from now on, become scalars
in four dimensions. These fields transform under $R$ as a vector
$v$, i.e.
\begin{eqnarray}
S &\supset& R \nonumber \\ adjS &=& adjR+v .
\end{eqnarray}
Moreover $\phi_a(x,y)$ acts as an intertwining operator connecting
induced representations of $R$ acting on $G$ and $S/R$. This
implies, exploiting Schur's lemma, that the transformation
properties of the fields $\phi_a(x,y)$ under $H$ can be found if
we express the adjoint representation of $G$ in terms of $R_{G}
\times H$ :
\begin{eqnarray}
G &\supset& R_{G} \times H \nonumber \\
 adjG &=&(adjR,1)+(1,adjH)+\sum(r_{i},h_{i}).
\end{eqnarray}
Then if $v=\sum s_{i}$, where each $s_{i}$ is an irreducible
representation of $R$, there survives an $h_{i}$ multiplet for
every pair $(r_{i},s_{i})$, where $r_{i}$ and $s_{i}$ are
identical irreps. of $R$. If we start from a pure gauge theory on
$M^4\times S/R$, the four-dimensional potential (at $y^a=0$) can
be shown to be given by
\begin{equation}\label{4-dimpot}
V= \frac{1}{4} F_{ab}F^{ab}= \frac{1}{4} (C^{\hat{c}}_{\
ab}\phi_{\hat{c}} - [\phi_{a}, \phi_{b}])^{2},
\end{equation}
where we have defined  $\phi_{i}\equiv W_{i}$. However, the fields
$\phi_a$ are not independent because the conditions
(\ref{condCSDR}) at $y^a=0$ constrain them. The solution of the
constraints provides the physical dimensionally reduced fields in
four dimensions; in terms of these physical fields the potential
is still a quartic polynomial. Then, the minimum of this potential
will determine the spontaneous symmetry breaking pattern.

Turning next to the fermion fields, similarly to scalars, they act
as an intertwining operator connecting induced representations of
$R$ in $G$ and in $SO(d)$, where $d$ is the dimension of the
tangent space of $S/R$. Proceeding along similar lines as in the
case of scalars, and considering the more interesting case of even
dimensions, we impose first the Weyl condition. Then to obtain the
representation of $H$ under which the four-dimensional fermions
transform, we have to decompose the fermion representation
$\rho_F$ of the initial gauge group $G$ under $R_{G} \times H$,
i.e.
\begin{equation}\label{26}
\rho_F= \sum (t_{i},h_{i}),
\end{equation}
and the spinor of $SO(d)$ under $R$
\begin{equation}\label{27}
\sigma_{d} = \sum \sigma_{j}.
\end{equation}
Then for each pair $t_{i}$ and $\sigma_{i}$, where $t_{i}$ and
$\sigma_{i}$ are identical irreps. there is an $h_{i}$ multiplet
of spinor fields in the four-dimensional theory. In order however
to obtain chiral fermions in the effective theory we may  have to
impose further requirements \cite{Kapetanakis:hf,Chapline:wy}.

\sk
\subsection{CSDR over fuzzy coset spaces}
In the present case space-time has the form $M^{4} \times
(S/R)_{F}$, where $(S/R)_{F}$ is the approximation of  $S/R$ by
finite $n\times n$ matrices. On  $M^{4} \times (S/R)_{F}$ we
consider a non-commutative gauge theory with gauge group $G=U(P)$.
We implement the CSDR scheme in the fuzzy case in three steps:\\
{\bf 1}) We state the CSDR principle on fuzzy cosets and reduce it to
a set of contraints --the CSDR constraints 
(\ref{3.17}), (\ref{3.19}), (\ref{eq7}), (\ref{eq16}), (\ref{eq16bis})-- 
that the gauge and matter fields must satisfy.\\
{\bf 2}) We reinterpret actions on $M^4\times (S/R)_F$
with $G=U(P)$ gauge group as actions on $M^4$
with $U(nP)$ gauge group.
More explicitely, we expand 
the fields on $M^4\times (S/R)_F$ in Kaluza-Klein modes on  
$(S/R)_F$. Since the algebra of functions on $(S/R)_F$ is finite 
dimensional we obtain a finite tower of modes; since $(S/R)_F$
is described by $n\times n$ matrices a basis for this mode expansion
is given by the generators of Lie$(U(n))$. In this way we show that the 
different modes can be conveniently grouped togheter so that
an initial Lie$(G)$-valued  field on $M^4\times (S/R)_F$ 
(with $G=U(P)$)
is reinterpreted as a Lie$(U(nP))$ valued field on  $M^4$.
Of course also the CSDR constraints can now be interpreted on $M^4$
instead of on $M^4\times (S/R)_F$. This leads to their solution in
step 3).\\
{\bf 3}) We solve the CSDR constraints and obtain
the gauge group and the particle content of the reduced
four-dimensional actions. This last step is first studied
in the fuzzy sphere case, and then for more general fuzzy 
cosets.

\subsubsection{CSDR principle}

\noindent 
Since the Lie algebra of $S$ acts on the fuzzy space $(S/R)_{F}$,
we can state the CSDR principle in the same way as in the
continuum case, i.e. the fields in the theory must be invariant
under the infinitesimal $S$ action up to an infinitesimal gauge
transformation
\begin{equation}
{\cal L}_{\hat{b}} \phi =\delta^{W_\b}\phi= W_{\hat{b}} \phi
~~~~~,~~~~~~ {\cal L}_{\hat{b}}A = \delta^{W_\b}A=-DW_{\hat{b}},
\label{csdr}
\end{equation}
where $A$ is the one-form gauge potential $A = A_{\mu}dx^{\mu} +
A_{\hat{a}} \theta^{\hat{a}}$, and $W_\b$ depends only on the
coset coordinates $X^\a$ and (like $A_\mu, A_a$) is antihermitian.
We thus write $W_\b=W_\b^\alpha{\cal T}^\alpha, \,\alpha=1,2\ldots
P^2,$ where ${\cal  T}^i$ are hermitian generators of $U(P)$ and
$(W_b^i)^\dagger=-W_b^i$, here ${}^\dagger$ is hermitian
coniugation on the $X^\a$'s. The principle gives for the
space-time part $A_{\mu}$
\begin{equation}\label{spacetime}
{\cal L}_{\hat b}A_{\mu} = [X_{\hat{a}} , A_{\mu}] = -[A_{\mu},
W_{\hat{b}} ],
\end{equation}
while for the internal part $A_{\hat a}$
\begin{equation}\label{internal}
[X_{\hat{b}}, A_{\hat{d}}] + A_{\hat{a}}C_{\hat{b}\ \ \hat{d}}^{\
\hat{a}} = - [A_{\hat{d}}, W_{\hat{b}}] - {\cal
L}_{\hat{d}}W_{\hat{b}}.
\end{equation}
From the first of eqs.~(\ref{csdr}) we have ${\cal L}_{\hat{a}}
{\cal L}_{\hat{b}} \phi = ({\cal L}_{\hat{a}}W_{\hat{b}}) \phi +
W_{\hat{b}}W_{\hat{a}}\phi $, then using the relation
$[{\L}_a,{\L}_\b]=C_{\a\b}^{~~\c}{\L}_\c$ we obtain the consistency
condition
\begin{equation}\label{2.59}
[X_{\hat{a}}, W_{\hat{b}}]-[X_{\hat{b}}, W_{\hat{a}}] -
[W_{\hat{a}}, W_{\hat{b}}] = C_{\hat{a} \hat{b}}^{\ \
\hat{c}}W_{\hat{c}}.
\end{equation}
Under the gauge transformation $\phi\to \phi'=g\phi$ with
$g \in G= U(P)$, we have ${\L}_\a\phi'=W'_\a\phi'$ and also
${\L}_\a\phi'= ({\L}_{\hat{a}}g) \phi + g ( {\L}_{\hat{a}}\phi)$, and
therefore
\begin{equation}\label{2.60}
W_\a\to W'_{\hat{a}} = gW_{\hat{a}}g^{-1} + [X_{\hat{a}},
g]g^{-1}~.
\end{equation}
Now in order to solve the constraints (\ref{spacetime}),
(\ref{internal}), (\ref{2.59}) we cannot follow the strategy
adopted in the commutative case where the constraints were studied
just at one point of the coset (say $y^a=0$). This is due to the
intrinsic nonlocality of the constraints. On the other hand the
specific properties of the fuzzy case (e.g. the fact that partial
derivatives are realized via commutators, the concept of covariant
derivative) allow to symplify and eventually solve the
constraints. If we define
\begin{equation}
\omega_{\hat{a}} \equiv X_{\hat{a}} - W_{\hat{a}}~,
\end{equation}
we obtain the following form of the consistency condition
(\ref{2.59})
\begin{equation}\label{3.17}
[ \omega_{\hat{a}} , \omega_{\hat{b}}] = C_{\hat{a} \hat{b}}^{\ \
\hat{c}} \omega_{c},
\end{equation}
where $\omega_{\hat{a}}$ transforms as
\begin{equation}
\omega_\a\to \omega'_{\hat{a}} = g\omega_{\hat{a}}g^{-1}.
\end{equation}
Now eq.~(\ref{spacetime})  reads
\begin{equation}\label{3.19}
[\omega_{\hat{b}}, A_{\mu}] =0.
\end{equation}
Furthermore by considering the covariant coordinate,
\begin{equation}\label{eq17}
\vphi_{\hat{d}} \equiv X_{\hat{d}} + A_{\hat{d}} ~
\end{equation}
we have \eq \vphi\to\vphi'=g\vphi g^{-1} \label{eq18}
\en
and eq.~(\ref{internal}) simplifies to
\begin{equation}\label{eq7}
C_{\hat{b} \hat{d} \hat{e}} \vphi^{\hat{e}} = [\omega_{\hat{b}},
\vphi_{\hat{d}} ].
\end{equation}
Therefore eqs. (\ref{3.17}) (\ref{3.19}) (\ref{eq7}) are the
constraints to be solved. Note that eqs.~(\ref{eq18}) and
(\ref{eq7}) have the symmetry
\begin{equation}\label{groundstate}
\vphi_{\hat{a}} \to \vphi_{\hat{a}} + \omega_{\hat{a}} ~,
\end{equation}
suggesting that $\omega_{\hat{a}}$ is a ground state around  which
we calculate the fluctuations $\vphi_{\hat{a}}$, and indeed, as
formula (\ref{pot}) for the potential shows, $\vphi_{\hat{a}}
=\omega_{\hat{a}}$ minimize the potential; in fact the potential
vanishes for this value of $\vphi_\a$.

\sk

One proceeds in a similar way for the spinor fields. The CSDR
principle relates the Lie derivative on a spinor $\psi$, that we
consider in the adjoint representation of $G$, to a gauge
transformation; recalling eqs. (\ref{definitions}) and
(\ref{spinlie11}) we have
\begin{equation}\label{eq15}
[X_{\hat{a}}, \psi] + \frac{1}{2}C_{\hat{a} \hat{b}
\hat{c}}\Gamma^{\hat{b} \hat{c}}\psi = [W_{\hat{a}},\psi],
\end{equation}
where $\psi$ denotes the column vector with entries $\psi_\alpha$.
Setting again $\omega_{\hat{a}} = X_{\hat{a}} - W_{\hat{a}}$ we
obtain the constraint
\begin{equation}\label{eq16}
-\frac{1}{2}C_{\hat{a} \hat{b} \hat{c}} \Gamma^{\hat{b}
\hat{c}}\psi = [\omega_{\hat{a}},\psi]~.
\end{equation}
We can also consider spinors that transform in the fundamental
rep. of  the gauge group $G$, we then have $[X_{\hat{a}}, \psi] +
\frac{1}{2}C_{\hat{a} \hat{b} \hat{c}}\Gamma^{\hat{b} \hat{c}}\psi
= W_{\hat{a}}\psi,
$
and setting again $\omega_{\hat{a}} = X_{\hat{a}} - W_{\hat{a}}$
we obtain
\begin{equation}\label{eq16bis}
-\frac{1}{2}C_{\hat{a} \hat{b} \hat{c}} \Gamma^{\hat{b}
\hat{c}}\psi = \omega_{\hat{a}}\psi-\psi X_\a~.
\end{equation}

\subsubsection{Actions and Kaluza-Klein modes }

\noindent
Let us consider a pure YM action on $M^{4} \times (S/R)_{F}$
and examine how it is reinterpreted in four dimensions. The action is
\begin{equation}
{\cal A}_{YM}={1\over 4} \int d^{4}x\, Tr\, tr_{G}\, F_{MN}F^{MN},
\end{equation}
where $Tr$ is the usual trace over $n\times n$ matrices and is
actually the integral over the fuzzy coset
$(S/R)_F\,$\footnote{$\,Tr$ is a good integral because it has the
cyclic property $Tr(f_1\ldots f_{p-1}f_p)=Tr(f_pf_1\ldots
f_{p-1})$. It is also invariant under the action of the group
$S$, that we recall to be infinitesimally given by ${\cal L}_{\hat{a}}
f = [{X}_{\hat{a}}, f ].$},
while $tr_G$ is the gauge group
$G$ trace. The higher-dimensional field strength $F_{MN}$
decomposed in four-dimensional space-time and extra-dimensional
components reads as follows
\begin{equation}
(F_{\mu \nu}, F_{\mu \hat{b}}, F_{\hat{a} \hat{b} } )~;
\label{decomp}
\end{equation}

\def\phi{\vphi}
\noindent explicitly the various components of the field strength
are given by
\begin{eqnarray}
F_{\mu \nu} &=&
\partial_{\mu}A_{\nu} -
\partial_{\nu}A_{\mu} + [A_{\mu}, A_{\nu}],\\[.3 em]
F_{\mu \hat{a}} &=&
\partial_{\mu}A_{\hat{a}} - [X_{\hat{a}}, A_{\mu}] + [A_{\mu},
A_{\hat{a}}] \nonumber\\[.3 em]
&=&
\partial_{\mu}\vphi_{\hat{a}} + [A_{\mu}, \vphi_{\hat{a}}] =
D_{\mu}\vphi_{\hat{a}},\\[.3 em] F_{\hat{a} \hat{b}} &=&
[\phi_{\hat{a}}, \phi_{\hat{b}}] - C^{\hat{c}}_{\ \hat{a} \hat{b}}
\phi_{\hat{c}}~;
\end{eqnarray}
they are covariant under {local} $G$ transformations:
$F_{MN}\to gF_{MN}g^{-1}$, with $g=g(x^\mu,X^\a)$.

In terms of the decomposition (\ref{decomp}) the action reads
\begin{equation}
{\cal A}_{YM}= \int d^{4}x\, Tr\, tr_{G}\,\left( {1\over 4}F_{\mu
\nu}^2+ {1\over 2}(D_{\mu}\phi_{\hat{a}})^2\right) - V(\phi)~,
\label{theYMaction}
\end{equation}
where the potential term $V(\phi)$ is the $F_{\hat{a} \hat{b}}$
kinetic term (recall $F_{\a\b}$ is antihermitian so that $V(\phi)$
is hermitian and non-negative)
\begin{eqnarray}\label{pot}
V(\phi)&=&-{1\over 4} Tr\,tr_G \sum_{\hat{a} \hat{b}} F_{\hat{a}
\hat{b}} F_{\hat{a} \hat{b}} \nonumber \\ &=& -{1\over 4}Tr\,tr_G
\sum_{\hat{a} \hat{b}} \left( [\phi_{\hat{a}}, \phi_{\hat{b}}] -
C^{\hat{c}}_{\ \hat{a} \hat{b}}
\phi_{\hat{c}}\right)\left([\phi_{\hat{a}}, \phi_{\hat{b}}] -
C^{\hat{c}}_{\ \hat{a} \hat{b}} \phi_{\hat{c}}\right)
\end{eqnarray}
{}For sake of clarity we here recall that: $Tr$ is the trace over the
$n\times n$ matrices that describe the fuzzy coset $(S/R)_F$, $tr_G$
is the trace over $G=U(P)$ matrices in the fundamental representation,
$\vphi$ is the covariant coordinate [cf. (\ref{eq17})]
where $X^\a$ is normalized as in (\ref{cas2}),
$rC^\a_{~\b\c}$ are the $S$ structure constants.
\sk

The action (\ref{theYMaction}) is naturally interpreted as an action in four
dimensions.
 The infinitesimal $G$ gauge transformation with gauge
parameter $\lambda(x^\mu,X^\a)$ can indeed be interpreted just as
an $M^4$ gauge transformation. We write 
\eq
\lambda(x^\mu,X^\a)=\lambda^\alpha(x^\mu,X^\a){\cal T}^\alpha
=\lambda^{h, \al}(x^\mu)T^h{\cal T}^\alpha~,\label{3.33}
\en
where ${\cal T}^\alpha$ are hermitian generators of $U(P)$,
$\lambda^\alpha(x^\mu,X^\a)$ are $n\times n$ antihermitian
matrices
and thus are expressible as $\lambda(x^\mu)^{\al , h}T^h$, where $T^h$
are antihermitian generators of $U(n)$. The fields 
$\lambda(x^\mu)^{\al , h}$, with $h=1,\ldots n^2$, are 
the Kaluza-Klein modes of $\lambda(x^\mu, X^\a)^{\al}$. 
We now consider on equal footing the indices $h$ and $\al$ 
and interpret the fields on the r.h.s. of (\ref{3.33})
as one field valued in the tensor product Lie algebra
${\rm{Lie}}(U(n)) \otimes {\rm{Lie}}(U(P))$. This Lie algebra 
is indeed ${\rm{Lie}}(U(nP))\,$\footnote{Proof: The $(nP)^2$
generators $T^h{\cal T}^\al$ are $nP\times nP$ antihermitian
matrices. We just have to show that they are linearly independent.
This is easy since it is equivalent to prove the linear
independence of the $(nP)^2$ matrices
$e_{ij}\varepsilon_{\rho\sigma}$ where $i=1,\ldots n$,
$\rho=1,\ldots P$ and $e_{ij}$ is the $n\times n$ matrix having
$1$ in the position $(i,j)$ and zero elswere, and similarly for
the $P\times P$ matrix $\varepsilon_{\rho\sigma}$.}. Similarly we
rewrite the gauge field $A_\nu$ as 
\eq
A_\nu(x^\mu,X^\a)=A_\nu^\alpha(x^\mu,X^\a){\cal T}^\alpha
=A_\nu^{h, \al}(x^\mu)T^h{\cal T}^\alpha~,
\en
and interpret it as a ${\rm{Lie}}(U(nP))$ valued gauge field on
$M^4$, and similarly for $\vphi_\a$. Finally $Tr\, tr_G$ is the trace
over $U(nP)$ matrices in the fundamental representation.
\sk

The above analysis applies also to more general actions, and to
the field $\omega_\a$ and therefore to
the CSDR constraints  (\ref{3.17}),
(\ref{3.19}), (\ref{eq7}), (\ref{eq16}), (\ref{eq16bis}) that 
can now be reinterpreted as
constraints on $M^4$ instead of on $M^4\times (S/R)_F$. 
The action
(\ref{theYMaction}) and the minima of the potential (\ref{pot}),
in the case $P=1$, have been studied, 
without CSDR constraints, in refs. \cite{DV.M.K., M.}.
It is imposing the CSDR constraints that we reduce the number of 
independent 
gauge and matter fields  in the action (\ref{theYMaction}), 
and that we therefore obtain new and richer particle models. 
We now solve these constraints.
We first consider the fuzzy sphere case and then extend the results
to more general fuzzy cosets.

\subsubsection{CSDR constraints for the fuzzy sphere}

\noindent We consider  $(S/R)_{F}=S^2_{F}$, i.e. the fuzzy sphere,
that we consider at fuzziness level $N$ (\,$(N+1) \times (N+1)$
matrices). We first study the basic example where the gauge group
$G$ is just $U(1)$. 
There we begin by considering a specific solution
--determined by a specific embedding of $SU(2)$ into $U(N+1)$-- of
constraint  (\ref{3.17}); we then solve also
(\ref{3.19}), (\ref{eq7}) and (\ref{eq16}), 
the latter concerns fermions in the adjoint 
of $G=U(1)$. We further write down the fermion 
action on $M^4\times S^2_F$, reinterpret it as an action on
$M^4$ (as we did for the pure YM action (\ref{theYMaction})), 
and  then rewrite the complete YM plus fermion action in terms of 
the fields that satisfy the CSDR constraints. 
We then describe in full generality 
how to solve the CSDR constraints  (\ref{3.17}), (\ref{3.19}), 
(\ref{eq7}) and (\ref{eq16}). Finally we study the case of
fermions that transform in the fundamental of the gauge group 
$G=U(1)$. The generalization of the above outlined
analysis to the case of the
gauge group $G=U(P)$ then follows. 
\sk

{\it The $G=U(1)$ case. }$~$
In this case the
$\omega_{\hat{a}}=\omega_{\hat{a}}(X^\b)$ that appear in the
consistency condition (\ref{3.17}), $[\omega_{\hat{a}},
\omega_{\hat{b}}] = C_{\hat{a} \hat{b}}^{\ \ \hat{c}}
\omega_{\hat{c}}$, are $(N+1) \times (N+1)$ antihermitian
matrices, i.e. we can interpret them as elements of
${\rm{Lie}}(U(N+1))$. On the other hand  $r \omega_{\hat{a}}$
satisfy the commutation relations (\ref{3.17}) of
${\rm{Lie}}(SU(2))$ (in fact $r C^a_{~bc}$ are the $SU(2)$ 
structure constants). Therefore in order to satisfy the consistency
condition (\ref{3.17}) we have to embed ${\rm{Lie}}(SU(2))$ in
${\rm{Lie}}(U(N+1))$. Let $T^h$ with $h = 1, \ldots ,(N+1)^{2}$ be
the generators of ${\rm{Lie}}(U(N+1))$ in the fundamental
representation and with normalization 
$Tr(T^hT^k)=-{1\over 2}\delta^{hk}$.
We can always use the convention $h= (\hat{a} ,
u)$ with $\hat{a} = 1,2,3$ and $u= 4,5,\ldots, (N+1)^{2}$ where
the $T^\a$ satisfy the $SU(2)$ Lie algebra,
\begin{equation}
[T^{\hat{a}}, T^{\hat{b}}] = r  C^{\hat{a} \hat{b}}_{\ \
\hat{c}}T^{\hat{c}}~.
\end{equation}
Then we define an embedding by identifying 
\eq
 r\omega_{\hat{a}}= T_{\hat{a}}.
\label{embedding}
\en
Constraint (\ref{3.19}), $[\omega_{\hat{b}} , A_{\mu}] = 0$, then
implies that the four-dimensional gauge group $K$ is the
centralizer of the image of $SU(2)$ in $U(N+1)$, i.e. $$
K=C_{U(N+1)}(SU((2))) = SU(N-1) \times U(1)\times U(1)~, $$ here
the last $U(1)$ is the $U(1)$ of $U(N+1)\simeq SU(N+1)\times
U(1)$. The functions $A_{\mu}(x,X)$ are arbitrary functions of $x$
but the $X$ dependence is such that $A_{\mu}(x,X)$ is
${\rm{Lie}}(K)$ valued instead of ${\rm{Lie}}(U(N+1))$, i.e.
eventually we have a four-dimensional gauge potential $A_\mu(x)$
with values in ${\rm{Lie}}(K)$. Concerning constraint
(\ref{eq7}), $[\omega_{\hat{b}} , \phi_{\hat{a}}] = C_{\hat{b}
\hat{a}}^{\ \ \hat{e}} \phi_{\hat{e}}$, we note that it is
satisfied by choosing 
\eq \label{soleasy} \vphi_\a=\vphi(x)
r \omega_\a~,
\en
i.e. the unconstrained degrees of freedom correspond to the scalar
field $\vphi(x)$ that is a singlet under the four-dimensional
gauge group $K$. This solution is unique since, given the
embedding (\ref{embedding}), the adjoint of $SU(2)$ is contained
just once in the adjoint of $U(N+1)$ (see for example
\cite{Fuchs:jv}). 

The physical spinor fields are obtained by solving
the constraint (\ref{eq16}), $-\frac{1}{2}C_{\hat{a} \hat{b}
\hat{c}} \Gamma^{\hat{b} \hat{c}}\psi = [\omega_{\hat{a}},\psi]$.
In the l.h.s. of this formula we can say that we have an embedding
of ${\rm{Lie}}(SU(2))$ in the spin representation of
${\rm{Lie}}(SO(3))$. This embedding is given by the matrices
$\tau^\a= \frac{1}{2}C_{\hat{a} \hat{b} \hat{c}} \Gamma^{\hat{b}
\hat{c}}$; since ${\rm{Lie}}(SU(2))= {\rm{Lie}}(SO(3))$ this
embedding is rather trivial and indeed $\tau^\a={-i\over 2
r}\sigma^\a$. Thus the constraint (\ref{eq16}) states that the
spinor $\psi=\psi^h T^h=
\left(^{}_{}\right.\!{^{_{_{{\mbox{$^{\psi_1}$}}}}}}_{^{^{{\mbox{$_{\!\!\!\!\!\!\!\!\psi_2}$}}}}}\!\left)^{}_{}\right.$,
where $T^h\in {\rm{Lie}}(U(N+1))$ and
$\psi_{1(2)}=\psi_{1(2)}^hT^h$ are four-dimensional spinors,
relate (intertwine) the fundamental rep. of $SU(2)$ to the
representations of $SU(2)$  induced by the embedding
(\ref{embedding}) of $SU(2)$ in $U(N+1)$, i.e. of $SU(2)$ in
$SU(N+1)$. In formulae
\begin{eqnarray}
SU(N+1) &\supset& SU(2) \times SU(N-1) \times U(1) \nonumber \\
(N+1)^{2}-1 &=& (1,1)_{0} \oplus (3,1)_{0} \oplus (1,
(N-1)^{2})_{0} \nonumber
\\ & & \oplus \, (2,(N-1))_{-(N+1)} \oplus (2,\overline{(N-1)})_{N+1}.
\end{eqnarray}
Then we deduce that the fermions that satisfy constraint
(\ref{eq16}) transform as $(N-1)_{-(N+1), 0}$ and
$\overline{(N-1)}_{N+1, 0}$ under $K= SU(N-1) \times U(1)\times
U(1)$. In the case of the fuzzy sphere the embedding
${\rm{Lie}}(SU(2)) \subset {\rm{Lie}}(SO(3))$ is somehow trivial.
If we had chosen instead the fuzzy $(SU(3)/U(1)\times U(1))_F$,
then ${\rm{Lie}}(SU(3))$ should be embedded in
${\rm{Lie}}(SO(8))$.
\sk

In order to write the action for fermions we have to
consider the Dirac operator ${\cal D}$ on $M^4\times S^2_F$. This
operator can be constructed following the derivation presented in
ref. \cite{Grosse:1995jt} for the Dirac operator on the fuzzy
sphere, see also ref. \cite{Ydri:2001pv}. {}For fermions in the
adjoint we obtain \eq {\cal
D}\psi=i\Gamma^\mu(\partial_\mu+A_\mu)\psi +i\sigma^\a
[X_{\hat{a}}+A_\a,\psi] - {1\over r}\psi~,
\en
where $\Gamma^\mu$ is defined in (\ref{Gamma7}), and with slight
abuse of notation we have written $\sigma^\a$ instead of
$\sigma^\a\otimes 1$. Using eq.~(\ref{eq17})
the fermion action,
\begin{equation}\label{fermionact}
{\cal A}_{F} =\int d^{4}x\, Tr \, \bar{\psi}{\cal D}\psi
\end{equation}
becomes
\begin{equation}
{\cal A}_{F} = \int\! d^{4}x\,\,\, Tr \,
\bar{\psi}\left(i\Gamma^\mu(\partial_\mu+A_\mu) -{1\over
r}\right)\psi\,+i\,
Tr \, \bar{\psi}\sigma^{\hat{a}} [\phi_{\hat{a}}  , \psi] \,
~,\label{341}
\end{equation}
where we recognize the fermion masses $1/r$ and the Yukawa
interactions. 

\sk\sk\sk
Using  eqs.~(\ref{soleasy}), (\ref{eq16}) 
the YM action (\ref{theYMaction}) plus the fermion action reads
\begin{eqnarray}
 {\cal A}_{YM}+{\cal A}_{F} &=& 
\int\! d^{4}x\,\,\, {1\over 4}Tr(F_{\mu
\nu}F^{\mu\nu})-{3\over 4}D_{\mu}\phi D^\mu\phi -
{3\over 8}(\phi^2-r^{-1}\phi)^2\,\nonumber\\
& &+
\int\! d^{4}x\,\,\, Tr \,
\bar{\psi}\left(i\Gamma^\mu(\partial_\mu+A_\mu) -{1\over
r}\right)\psi\,-\, {3\over 2 }\,Tr \, \bar{\psi} \phi \psi \, ~.
\end{eqnarray}
\sk

The choice (\ref{embedding}) defines one of the possible embeddings
of ${\rm{Lie}}(SU(2))$ in ${\rm{Lie}}(U(N+1))$
[${\rm{Lie}}(SU(2))$ is embedded in ${\rm{Lie}}(U(N+1))$ as a
regular subalgebra], while on the other extreme we can embed
${\rm{Lie}}(SU(2))$ in ${\rm{Lie}}(U(N+1))$ using the irreducible
$N+1$ dimensional rep. of $SU(2)$, i.e. we identify
$\omega_{\hat{a}}= X_{\hat{a}}$. Constraint (\ref{3.19}) in this
case implies that the four-dimensional gauge group is $U(1)$ so
that $A_\mu(x)$ is $U(1)$ valued. Constraint  (\ref{eq7}) leads
again to the scalar singlet $\vphi(x)$.

In general, we start with a $U(1)$ gauge theory on $M^4\times
S^2_F$. We solve the CSDR constraint (\ref{3.17}) by embedding
$SU(2)$ in $U(N+1)$. There are $p(N+1)$ embeddings where $p(n)$ is
the number of ways one can partition the integer $n$ into a set of
non-increasing positive integers \cite{Mad} (for example the
solution $\omega^\a=0$ corresponds to the partition $(1,1,\ldots
1)$, and the embedding using the $n$ irrep. of $SU(2)$ corresponds
to the partition $(n)\,$). Then  constraint (\ref{3.19}) gives the
surviving four-dimensional gauge group. Constraint (\ref{eq7})
gives the surviving four-dimensional scalars and eq.
(\ref{soleasy}) is always a solution but in general not the only
one. Setting $\phi_\a=\omega_\a$ we always minimize the potential.
This minimum is given by the chosen embedding of $SU(2)$ in
$U(N+1)$. Constraint (\ref{eq16}) gives the surviving four
dimensional spinors. \sk

{}Finally  let us consider spinors that transform in the
fundamental of the gauge group $G$. Then in the fermion action
(\ref{fermionact}) we have the covariant Dirac operator
$
{\cal D}\psi=i\Gamma^\mu(\partial_\mu+A_\mu)\psi +i\sigma^\a
[X_{\hat{a}},\psi] +i\sigma^\a A_\a\psi - {1\over r}\psi~,
$
and instead of constraint (\ref{eq16}) we have to use constraint
(\ref{eq16bis}). We thus obtain \eq {\cal A}_{F} = \int\!
d^{4}x\,\,\, Tr \, \bar{\psi}\left(i\Gamma^\mu(\partial_\mu+A_\mu)
-({5\over 2 r}+i\sigma^\a\omega_\a)\right)\psi\,+i\, Tr \,
\bar{\psi}\sigma^{\hat{a}} \phi_{\hat{a}} \psi \, ~.
\en
In the following we study constraint (\ref{eq16bis}) in the
example where constraint (\ref{3.17}) is solved by considering the
embedding \eq \label{embe} SU(N+1)\supset SU(2) \times U(1)~,
\en
obtained by identifying $\omega_\a$ with the generators of $SU(2)$
in the $N$ dimensional irrep.. This embedding induces the
embedding and the branching rule\footnote{The generator
$\lambda=diag(1,1,\ldots 1,-N)$ of the first $U(1)$ appearing in
the r.h.s. of (\ref{emmbedd}) is normalized so that
$Tr(\lambda^2)=N(N+1)$. This implies the normalization
$\lambda'=\sqrt{N}diag(1,1,\ldots 1,1)$ for the generator of the
second $U(1)$ appearing in the r.h.s. of  (\ref{emmbedd}), i.e.
the $U(1)$ coming from $U(N+1)\simeq SU(N+1)\times U(1)$.}
\begin{eqnarray}
\label{emmbedd} U(N+1)\;\simeq\; SU(N+1)\times U(1)&\supset& SU(2)
\times U(1)\times U(1)~,\\
(N+1)_{\sqrt{N}}&=&N_{1,\sqrt{N}\,}\oplus
1_{-N,\sqrt{N}}\nonumber~~~.
\end{eqnarray}
It follows that $SU(2)$ acts on $\psi$ via the representation
$(N\oplus 1)\times \overline{(N+1)}$ given by $\delta\psi=
\omega_\a\psi-\psi X_\a$, cf. (\ref{eq16bis}). We have \eq
\label{rrep} (N_{1,\sqrt{N}\,}\oplus 1_{-N,\sqrt{N}})\times
\overline{(N+1)} =(N_{1,\sqrt{N}\,}\oplus 1_{-N,\sqrt{N}})\times
{(N+1)}~~~~~~~~~~~~~~~~~~~~~~~~~~~~~~~~~~~~\vspace{-.2cm}
\en
$$
~~~~~~~~~~~~~~~~~~~~~~~~~=2N_{1,\sqrt{N}\,}\oplus(2N-2)_{1,\sqrt{N}\,}\oplus
(2N-4)_{1,\sqrt{N}\,}\ldots \oplus 2_{1,\sqrt{N}} \oplus
(N+1)_{-N,\sqrt{N}}~, \nonumber $$ where the indices denote the
eigenvalues of the $U(1)$ generators appearing in the r.h.s. of
(\ref{emmbedd}). We can now solve constraint (\ref{eq16bis}) that
states that the spinor $\psi$  intertwines the fundamental rep. of
$SU(2)$ appearing in the l.h.s. of (\ref{eq16bis}) with the rep.
of $SU(2)$ appearing in the r.h.s. of (\ref{eq16bis}). Since this
latter in the present  example contains the 2 of $SU(2)$ just
once, we conclude that there exists one surviving four-dimensional
spinor; this spinor has charges $(1,\sqrt{N})$ with respect to the
four-dimensional gauge group $K=U(1)\times U(1)$. In general for
fermions in the fundamental we consider the product of the
$\overline{(N+1)}$ of $SU(2)$ times the representations of $SU(2)$
on $\psi$ induced by the embedding of $SU(2)$ in $U(N+1)$ [in the
above example the embedding defined by (\ref{emmbedd})]. There are
as many four-dimensional spinors as many times the fundamental of
$SU(2)$ appears in this product. \sk

\sk
{\it The $G=U(P)$ case. } In this case
$\omega_{\hat{a}}=\omega_{\hat{a}}(X^\b)=\omega_\a^{h,\al}T^h{\cal
T}^\al$ is an $(N+1)P\times (N+1)P$ hermitian matrix and in order
to solve the constraint (\ref{3.17}) we have to embed
${\rm{Lie}}(SU(2))$ in  ${\rm{Lie}}(U((N+1)P))$. All the results
of the $G=U(1)$ case holds also here, we just have to replace
$N+1$ with $(N+1)P$. This is true for the fermion sector too,
provided that in the higher dimensional theory the fermions are
considered in the adjoint of $U(P)$ (in the action
(\ref{fermionact}) we then need to replace $Tr$ with $Tr_{\,}
tr_{U(P)}$ i.e. $tr_{U((N+1)P)}$). We can also consider fermions in
the fundamental of $U(P)$. Then an infinitesimal gauge
transformation reads $\delta\psi=\lambda\psi$ and the
four-dimensional spinors $\psi_{1(2)}$, where
$\psi=\left(^{}_{}\right.\!{^{_{_{{\mbox{$^{\psi_1}$}}}}}}_{^{^{{\mbox{$_{\!\!\!\!\!\!\!\!\psi_2}$}}}}}\!\left)^{}_{}\right.$,
transforms according to the fundamental of $U(P)$ and the $n\times
\overline{n}$ of $U(n)$, i.e. they transform according to the
fundamental of $U(nP)$ and the antifundamental of $U(n)$ (where
$n=N+1$). In this
case, in order to solve constraint (\ref{eq16bis}) and find the
surviving four-dimensional spinors, we have to consider the
product of the $SU(2)$ representation $\overline{(N+1)}=(N+1)$
times the representations  of $SU(2)$ on $\psi$ induced by the
embedding of $SU(2)$ in $U((N+1)P)$. The $SU(2)$ representation
$\overline{(N+1)}$ arises from the $SU(2)$ action
$\delta\psi=-\psi X_\a$ observing that $X_\a$ is an $SU(2)$
generator in the  irrep. $N+1$. There are as many four-dimensional
spinors as many times the fundamental of $SU(2)$ appears in this
product of representations.

\subsubsection{CSDR constraints for fuzzy cosets}

\noindent Consider a fuzzy coset $(S/R)_F$ (e.g. fuzzy $CP^M$)
described by $n\times n$ matrices, and let the higher dimensional
theory have gauge group $U(P)$. Then we see that constraint
(\ref{3.17}) implies that we have to embed $S$ in $U(nP)$.
Constraint (\ref{3.19}) then implies that the four dimensional
gauge group $K$ is the centralizer of the image $S_{U(nP)}$  of
$S$ in $U(nP)$, $K=C_{U(nP)}(S_{U(nP)})$.

Concerning fermions in the adjoint, in order to solve constraint
(\ref{eq16}) we consider the embedding $$S \subset SO(dimS)~, $$
which is given by $\tau_{\hat{a}} = \frac{1}{2}C_{ \hat{a} \hat{b}
\hat{c}} \Gamma^{\hat{b} \hat{c}}$ that satisfies
$[\tau^\a,\tau^b]=C_{\a\b\c}\tau^\c$. Therefore $\psi$ is an
intertwining operator between induced representations of $S$ in
$U(nP)$ and  in $SO(dimS)$. To find the surviving fermions, as in
the commutative case \cite{Kapetanakis:hf}, we decompose the
adjoint rep. of $U(nP)$ under $S_{U(nP)}\times K$,
\begin{eqnarray}
U(nP) &\supset& S_{U(nP)} \times K \nonumber \\ adjU(nP) &=&
\sum_{i} (s_i, k_i)~.
\end{eqnarray}
We also decompose the spinor rep. $\sigma$ of $SO(dimS)$ under $S$
\begin{eqnarray}
SO(dimS) &\supset& S \nonumber \\
 \sigma &=& \sum_{e} \sigma_{e}~.
\end{eqnarray}
Then, when we have two identical irreps. $s_i = \sigma_e$, there
is a $k_i$  multiplet of fermions surviving in four dimensions,
i.e. four-dimensional spinors $\psi(x)$ belonging to the $k_i$
representation of $K$.

Concerning fermions in the fundamental of the gauge group $U(P)$,
we recall that they can be interpreted as transforming according
to the fundamental of $U(nP)$ and the antifundamental
$\overline{n}$ of $U(n)$. Moreover the coordinates $X_\a$ are
generators of $S$ in the irrep. $n$, so that the $S$ action
$\delta\psi=-\psi X_\a$ is given by the irrep. $\overline{n}$. In
order to solve constraint (\ref{eq16bis}) we therefore  decompose
the fundamental of $U(nP)$ under $S_{U(nP)}\times K\,$, \eq
\,nP=\sum_i(t_i,h_i)~,
\en
 and then
consider the product representation \eq \sum_i(t_i\times
\overline{n}, h_i) =\sum_\ell(u_\ell,h_\ell)~,
\en
where now $u_\ell$ are irreps. of $S$. When we have two identical
irreps. $u_\ell = \sigma_e$, there is an $h_\ell$  multiplet of
fermions surviving in four dimensions, i.e. four-dimensional
spinors $\psi(x)$ belonging to the $h_\ell$  representation of
$K$.

\section{Discussion and Conclusions}
Non-commutative Geometry has been regarded as a promising
framework for obtaining finite quantum field theories and  for
regularizing quantum field theories. In general quantization of
field theories on non-commutative spaces has turned out to be much
more difficult and with less attractive ultraviolet features than
expected \cite{Filk:dm, Minwalla:1999px}, see however ref.
\cite{Grosse:2003nw}, and ref. \cite{Steinacker}, where pure
Yang-Mills theory on the fuzzy sphere is quantized. Recall also
that non-commutativity is not the only suggested tool for
constructing finite field theories. Indeed four-dimensional finite
gauge theories have been constructed in ordinary space-time and
not only those which are ${\cal N} = 4$ and ${\cal N} = 2$
supersymmetric, and most probably phenomenologically
uninteresting, but also chiral ${\cal N} = 1$ gauge theories
\cite{Kapetanakis:vx} which already have been successful in
predicting the top quark mass and have rich phenomenology that
could be tested in future colliders
\cite{Kapetanakis:vx,Kubo:1994bj}. In the present work we have not
adressed the finiteness of non-commutative quantum field theories,
rather we have used non-commutativity to produce, via Fuzzy-CSDR,
new particle models from  particle models on $M^4\times (S/R)_F$.

The Fuzzy-CSDR has different features from the ordinary CSDR
leading  therefore to new four-dimensional particle models. In
this paper we have established the rules for the construction of
these models; it may well be that Fuzzy-CSDR provides more
realistic four-dimensional theories. Having in mind the
construction of realistic models one can also combine the fuzzy
and the ordinary CSDR scheme, for example considering $M^4\times
S'/{R'}\times (S/R)_F$.

A major difference between fuzzy and ordinary SCDR is that in the
fuzzy case one always embeds $S$ in the gauge group $G$ instead of
embedding just $R$ in $G$. This is due to the fact that the
differential calculus used in the Fuzzy-CSDR is based on $dim S$
derivations instead of the restricted $dim S - dim R$ used in the
ordinary one.  As a result the four-dimensional gauge group $H =
C_G(R)$ appearing in the ordinary CSDR after the geometrical
breaking and before the spontaneous symmetry breaking due to the
four-dimensional Higgs fields does not appear in the Fuzzy-CSDR.
In Fuzzy-CSDR the spontaneous symmetry breaking mechanism takes
already place by solving the Fuzzy-CSDR constraints. The four
dimensional potential has the typical ``maxican hat'' shape,
but it appears already spontaneously broken. Therefore 
in four dimensions appears only the physical Higgs field
that survives after a spontaneous symmetry breaking.
Correspondingly in the Yukawa sector of the theory we have the
results of the spontaneous symmetry breaking, i.e. massive
fermions and Yukawa interactions among fermions and the physical
Higgs field. Having massive fermions in the final theory is a
generic feature of CSDR when $S$ is embedded in $G$ (see last ref.
in [20]). We see that if one would like to describe the
spontaneous symmetry breaking of the SM in the present framework,
then one would be naturally led to large extra dimensions.

A fundamental difference between the ordinary CSDR and its fuzzy
version is the fact that a non-abelian gauge group $G$ is not
really required in high dimensions. Indeed  the presence of a
$U(1)$ in the higher-dimensional theory is enough to obtain
non-abelian gauge theories in four dimensions. We plan to
elaborate further on this point, as well as on the possibility to
construct realistic theories.

\section*{Acknowledgments}
We would like to thank I.~Bakas, L.~Castellani, A.~Dimakis,
P.~Forgacs, H.~Grosse, B.~Jur{\v c}o, D.~Luest, L.~Palla and H.
Steinacker for useful discussions. G.~Z. would like to thank the
LAPTH in Annecy for the warm hospitality. P.~A. acknowledges
partial support by EU under the RTN contract HPRN-CT-2000-00131.
P.~M. and G.~Z. acknowledge partial support by EU under the RTN
contract HPRN-CT-2000-00148, the Greek-German Bilateral Programme
IKYDA-2001 and by  the NTUA programme for fundamental research
``THALES".


\end{document}